\documentclass[iop]{emulateapj}

\newcommand{\be}{\begin{equation}}
\newcommand{\ee}{\end{equation}} 

\shorttitle{High-$z$ Galaxies}
\shortauthors{Melia}

\begin{document}

\title{The Premature Formation of High Redshift Galaxies}

\author{Fulvio Melia\thanks{John Woodruff Simpson Fellow.}}
\affil{Department of Physics, The Applied Math Program, and Steward Observatory\\ 
The University of Arizona, AZ 85721, USA\\
\email{fmelia@email.arizona.edu}}

\begin{abstract}
Observations with WFC3/IR on the {\it Hubble} Space Telescope and the use of 
gravitational lensing techniques have facilitated the discovery of galaxies 
as far back as $z\sim 10-12$, a truly remarkable achievement. However, this 
rapid emergence of high-$z$ galaxies, barely $\sim 200$ Myr after the transition 
from Population III star formation to Population II, appears to be in conflict 
with the standard view of how the early Universe evolved. This problem has much 
in common with the better known (and probably related) premature appearance of 
supermassive black holes at $z\ga 6$. It is difficult to understand how $\sim 
10^9\;M_\odot$ black holes could have appeared so quickly after the big bang 
without invoking non-standard accretion physics and the formation of massive 
seeds, neither of which is seen in the local Universe. In earlier work, we 
showed that the appearance of high-$z$ quasars could instead be understood 
more reasonably in the context of the $R_{\rm h}=ct$ Universe, which does not 
suffer from the same time compression issues as $\Lambda$CDM does at early epochs. 
Here, we build on that work by demonstrating that the evolutionary growth of
primordial galaxies was consistent with the current view of how the first stars
formed, but only with the timeline afforded by the $R_{\rm h}=ct$ cosmology. We 
also show that the growth of high-$z$ quasars was mutually consistent with that 
of the earliest galaxies, though it is not yet clear whether the former grew 
from $5-20\;M_\odot$ seeds created in Population III or Population II supernova 
explosions.
\end{abstract}


\section{Introduction}
The first stars and galaxies started forming towards the end of the 
cosmic ``Dark Ages," a period thought to have lasted $\sim 400$ Myr
(in the context of $\Lambda$CDM) following hydrogen recombination 
at cosmic time $t\sim 0.4$ Myr. Over the past decade, many
detailed simulations of this process have produced a comprehensive 
picture of how the Universe transformed from a simple initial (dark) 
state to the complicated hierarchical system we see today, including
galaxy clusters and the many components contained within them 
(see, e.g., Barkana \& Loeb 2001;
Miralda-Escud\'e 2003; Bromm \& Larson 2004; Ciardi \& Ferrara 2005; 
Glover 2005; Greif et al. 2007; Wise \& Abel 2008; Salvaterra et al. 
2011; Greif et al. 2012; Jaacks et al. 2012; for more recent reviews, 
see also Bromm et al. 2009, and  Yoshida et al. 2012).

Interest in this cosmic dawn has been heightened recently with the
dramatic discovery of faint galaxies at redshifts well beyond the
end of the Epoc of Reionization (EoR), which observations show
started at the end of the Dark Ages ($z\sim 15$, i.e., $t\sim 400$ Myr)
and lasted until $z\sim 6$ ($t\sim 900$ Myr) (see, e.g., Zaroubi
2012). Stretching the imaging capabilities of WFC3/IR on the 
{\it Hubble} Space Telescope (HST), and using gravitational 
lensing techniques, investigators have apparently uncovered 
the earliest galaxies emerging as far back as $z\sim 10-12$, a 
truly remarkable achievement (Bouwens et al. 2011; Zheng et 
al. 2012; Ellis et al. 2012; Bouwens et al. 2012; Coe et al. 2013; 
Oesch et al. 2013; Brammer et al. 2013; Bouwens et al. 2013). 
These nascent galaxies may have contributed to the re-ionization 
of the intergalactic medium (IGM), perhaps even dominated
this process. 

But this rapid emergence 
of high-$z$ galaxies so soon after the big bang may actually be in 
conflict with our current understanding of how they came to be.  
This problem is very reminiscent of the better known (and probably 
related) premature appearance of supermassive black holes at 
$z\ga 6$. It is difficult to understand how 
$\sim$$10^9\;M_\odot$ black holes appeared so quickly after 
the big bang without invoking non-standard accretion physics 
and the formation of massive seeds, both of which are not 
seen in the local Universe. Recent observations (e.g., Jiang 
et al. 2007; Kurk et al. 2009; Willott et al. 2010; Mortlock et 
al. 2011; De Rosa et al. 2011) have compounded this problem 
by demonstrating that most (if not all) of the high-$z$ quasars 
appear to be accreting at their Eddington limit.  

In our recent assessment of the high-$z$ quasar problem
(Melia 2013a), we considered the possibility
that this conflict may be due to our use of an incorrect 
redshift-time relationship, i.e., an incorrect cosmological 
expansion, rather than to the astrophysics of black-hole 
formation and growth, which are increasingly constrained 
by the ever improving observations. We showed that the 
high-$z$ quasar data may instead be interpreted more 
sensibly in the context of the $R_{\rm h}=ct$ Universe 
(Melia 2007, 2012; Melia \& Shevchuk 2012), for which 
standard $5-20\;M_\odot$ seeds forming after re-ionization 
had begun at $z\la 15$ could have easily grown into $\ga 10^9
\;M_\odot$ supermassive black holes by redshift $z\ga 6$, 
merely by accreting at the observed Eddington rate. 
The principal difference between $\Lambda$CDM and 
$R_{\rm h}=ct$ that eases the tension between theory and 
observations is simply $z(t)$. Specifically, in the $R_{\rm h}=ct$ 
cosmology, the EoR began at $\sim 900$ Myr ($z\sim 15$ in this
cosmology) and ended at $\sim 1.9$ Gyr ($z\sim 6$), placing the 
birth of supermassive black holes at $\sim 1$ Gyr (roughly $z=13$), 
right where one would have expected them to form via the 
supernova deaths of Population III and II stars that presumably 
started the EoR.

Now that the premature emergence of high redshift galaxies is creating 
comparable tension with established theory, it is necessary to consider 
whether their timing issues are similarly resolved by the $R_{\rm h}=ct$ 
Universe, and whether one can find consistency between their 
evolutionary history and the growth of supermassive black 
holes at $z\ga 6$. In \S~2 of this paper, we will briefly review 
the current thinking behind the formation of structure at
$z\ga 6$, and then in \S~3 compare this with what the most 
recent observations are telling us. In \S~4, we will discuss 
a re-interpretation of high-$z$ galaxy growth in the context 
of $R_{\rm h}=ct$, and then present our conclusions in \S~5. 

\section{The Current Theoretical Picture}
\subsection{Background Physics}
Simulations tracing the growth of initial perturbations in the
expanding medium begin with a set of well-posed initial conditions,
derived from the fluctuation power spectrum and elemental abundances 
of the plasma constrained by observations of the cosmic microwave 
background (CMB; Komatsu et al. 2009). These calculations show that
the first stars (Population III) formed by redshift $z\sim 20$
at the core of dark matter minihaloes of mass $M_{\rm halo}\sim 
10^6\;M_\odot$ (Haiman et al. 1996; Tegmark et al.
1997; Abel et al. 2002; Bromm et al. 2002). In the concordance
$\Lambda$CDM model with parameter values
$\Omega_{\rm m}=0.27$, $\Omega=1$, $w_\Lambda=-1$,
and a Hubble constant $H_0\approx 70$ km s$^{-1}$ Mpc$^{-1}$,
this redshift corresponds to a cosmic time $t\approx 200$ Myr.
Here, $\Omega\equiv \rho/\rho_{\rm  c}$ is the density $\rho$ scaled 
to the critical (or closure) density $\rho_{\rm c}$ ($\equiv 3c^2H_0^2/
8\pi G$) in a flat universe, $\Omega_{\rm m}$ is the correspondingly
scaled (luminous and dark) matter density, and $w_\Lambda$ is
the dark energy (assumed to be a cosmological constant $\Lambda$) 
equation-of-state parameter yielding the pressure in terms of 
$\rho_\Lambda$.

It is difficult to see how Population III stars could have formed 
any earlier than this, since primordial gas in the early Universe 
could not cool radiatively. Only after molecular hydrogen started 
to accumulate could the plasma cool and eventually condense
to make stars (Galli \& Palla 1998). Once the density fluctuations 
had condensed to a minimum temperature of about $200$ K, set
by the internal rovibrational transitions of $H_2$ (Omukai
\& Nishi 1998), the gas cloud could become Jeans-unstable
and collapse to form a protostar at the center of the halo.
These objects, which would eventually become Population
III stars, attained a mass of $\ga 100\;M_\odot$ once
they reached the main sequence (Kroupa 2002; 
Chabrier 2003). Feedback from these first stars determined
the fate of the surrounding primordial gas clouds, because
the UV radiation from a single massive star could destroy all
of the $H_2$ in the parent condensation. As such, probably
only one Population III star could form in each minihalo
(Yoshida et al. 2008). These early structures were not
galaxies.

However, the enormous flux of ionizing radiation and 
$H_2$-dissociating Lyman-Werner radiation emitted by
massive Population III stars had an impact on more
than just their immediate surroundings (Bromm et al. 2001; 
Schaerer 2002); this radiation field apparently had
a dramatic influence out to several kiloparsecs or more
of the progenitor. In combination with strong outflows,
which lowered the density in the minihaloes, this ourpouring
of radiative and mechanical energy delayed subsequent
star formation considerably (Johnson et al. 2007).

The gas expelled by this first generation of Population III
stars would have been too hot and diffuse to permit further
star formation until it had time to cool and reach high densities
again. But cooling and re-collapse take a long time. Analytic
models (Yoshida et al. 2004), confirmed by detailed numerical
simulations (Johnson et al. 2007), show that the gas
re-incorporation time was roughly $100$ Myr, essentially the 
dynamical time for a first-galaxy halo to assemble itself.
The critical mass for hosting the formation of the first galaxies
is not yet known with high precision, though it is thought that
these would have probably been atomic cooling haloes, i.e., 
condensations with mass $\sim 10^8\;M_\odot$ and virial
temperatures $\ga 10^4$ K, so that atomic line cooling
could have been possible (Wise \& Abel 2007). This mass
scale is supported by simulations of the energetics, feedback, 
and chemical enrichment associated with the first (Population
III) supernova explosions (Greif et al. 2007). The supernova
remnant propagates for a Hubble time at $z\sim 20$ to a
final mass-weighted mean shock radius of 2.5 kpc, sweeping
up a total gas mass of $2.5\times 10^5\;M_\odot$ in the process.
A dark matter halo of at least $10^8\;M_\odot$ must be
assembled to recollect and mix all components of the shocked
gas. 

At $t\sim 300$ Myr, with $z\rightarrow 15$ towards the end of 
the Dark Ages, the formation of structure would have begun in 
earnest. But this is where a potential conflict emerges
between theoretical expectations and the recent observation 
of high-$z$ galaxies because, in $\Lambda$CDM, there simply 
wasn't sufficient time for these galaxies to form by $z\sim 
10-12$, a mere $\sim 200$ Myr later. We will demonstrate
this quantitatively in the next section, but first, let us see
what the simulations predict.

\subsection{A Brief Survey of the Simulations}
The results of simulations by independent workers essentially
confirm each other's conclusions, chiefly because their calculations
incorporate the same set of physical principles described above.
We will summarize the key features of two of these simulations,
by Jaacks et al. (2012) and Salvaterra et al. (2013), who use
different metrics to characterize their results, though with
very similar outcomes.

Starting their simulation at $z=100$, Salvaterra et al. (2013)
follow the growth of structure all the way out to $z=2.5$ and
find that by redshift $z\sim 6-10$, most of the early galaxies 
had a mass $\sim 10^6-10^8\;M_\odot$, with a few percent as high
as $\sim 10^9\;M_\odot$. These structures had a specific
(i.e., per solar-mass) star-formation rate (sSFR) $\sim 3-10$ 
Gyr$^{-1}$. The $10^8-10^9\;M_\odot$ galaxies were therefore
forming stars at a rate SFR $\sim 0.3-10\;M_\odot$ yr$^{-1}$,
in line with SFRs observed over a remarkably broad range of
redshifts ($z\sim 2-7$). In this redshift range, observations
indicate that galaxies of this size formed $2-3\;M_\odot$ per
year per $10^9\;M_\odot$ of total stellar mass (Stark et al.
2009; Gonzalez et al. 2010; McLure et al. 2011). As we shall
see in the next section, the recent discoveries appear to be
tracing the high end of this mass and SFR distribution (see 
Table 1).

The key result of this work most relevant to our discussion
in this paper is that the ratio between the doubling time
$t_{\rm db}$ (i.e., the inverse of the sSFR) and the corresponding
cosmic time seems to be universally equal to $\sim 0.1-0.3$,
independently of redshift. Let us conservatively take the
smaller value, which minimizes the growth time. Then, a galaxy 
with mass $M_*=10^8\;M_\odot$ at $z=6$ (corresponding
to $t_*\sim 900$ Myr in $\Lambda$CDM), would have started 
its condensation at cosmic time $t_{\rm init}\sim (0.9)^n\,
t_*$, where $n\equiv [\log (M_*/M_{\rm init})]/
\log 2$ is the number of doublings since the galaxy
started growing with a mass $M_{\rm init}$ at $t_{\rm init}$. 
Conservatively putting $M_{\rm init}=10^4\;M_\odot$, this
gives $t_{\rm init}\approx 230$ Myr, roughly where the
arguments concerning the transition from Population III
to Population II stars would placed it. 

In other words, these simulations appear to be consistent
with the growth implied by $10^8\;M_\odot$ galaxies 
at $z=6$. But let us now consider whether this same type
of growth rate would allow a similar galaxy to have formed
by $z=10$ (i.e., $t_*\approx 550$ Myr in $\Lambda$CDM). Such
a galaxy would have begun its growth with 
$M_{\rm init}=10^4\;M_\odot$ at $t_{\rm init}\sim 140$ Myr, 
well before even the Population III 
stars would have had sufficient time to develop and explode, 
producing the conditions necessary to initiate the subsequent 
growth of galactic structure. A $10^9\;M_\odot$ galaxy
observed at $z\sim 10.7$, corresponding to a cosmic time 
$t_*\sim 490$ Myr, would have started at $t_{\rm init}\sim 
82$ Myr, which appears to be completely untenable, given
what we now believe occurred at the dawn of cosmic structure
formation.

Jaacks et al. (2012) took a different approach and
examined the duty cycle and history of the SFR for high-redshift 
galaxies at $z\ga 6$, and found that, though individual galaxies 
have ``bursty" SFRs, the averaged SFR between $z\sim 15$ and 
$z\sim 6$ can be characterized well by an exponentially increasing
functional form with characteristic time-scales $t_{\rm c}
=70$ Myr to $200$ Myr, for galaxies with stellar mass
$M_*\sim 10^6\;M_\odot$ to $\sim 10^{10}\;M_\odot$,
respectively.

From their hydrodynamic simulations, one may therefore infer 
that stellar mass grew at a rate
\begin{equation}
{dM\over dt}=K\exp\left({t-t_*\over t_c}\right)\;,
\end{equation}
using the definition that a galaxy has mass $M_*$ at cosmic
time $t_*$. In these calculations, galaxies begin with a mass
$M<< M_*$ at $t\sim 400$ Myr and, for $M_*=10^8\;M_\odot$ at 
$t_*=900$ Myr (i.e., $z=6$), $t_c\sim 100$ Myr. Therefore,
$K\approx 1\;M_\odot$ yr$^{-1}$ in Equation~(1). Placing
a comparable $M=10^8\;M_\odot$ galaxy at $z\sim 10$, where 
$t\approx 550$ Myr, would then mean that its mass at 
$t=300$ Myr would have been $\sim 8\times 10^6\;M_\odot$,
which is inconsistent with the transition from Population
III to Population II stars described above.

Both of these calculations therefore suggest that,
whereas $10^8\;M_\odot$ galaxies would have had no problem 
forming by redshift 6 with the timeline predicted by 
$\Lambda$CDM, similar galaxies at $z\sim 10$ could not have 
grown to this mass during the short time elapsed since the
emergence of Population III and II stars. We are not aware
of any other simulations published thus far that produce
an outcome contrary to this result.

In the next section, we will summarize the observational status
of these high-$z$ galaxies, including a discussion of the actual,
measured SFR as a function of redshift. The empirically derived
expression for SFR(t) for galaxies at $z\la 8$ matches the theoretical 
predictions quite well. Therefore, when we use these relations
to trace comparably massive galaxies at larger redshifts, the 
results reinforce the view that $\sim 10^8-10^9\;M_\odot$ galaxies
probably could not have had sufficient time to form by $z\sim 10-12$
in $\Lambda$CDM.

\section{High Redshift Galaxies and their Evolutionary History}
\subsection{The High-$z$ Galaxy Sample}
Observations by {\it Spitzer} of the assembled stellar mass
at $z\sim 5-6$ suggest that star formation must have started
well beyond $z\sim 8$ (Stark et al. 2007; Gonzalez et al. 2010),
though only recently have attempts succeeded in finding the
earliest galaxies. Various groups have utilized an assortment
of techniques, some based on gravitational lensing by foreground
clusters of galaxies, others relying on spectral energy distribution
fitting techniques on objects selected from a deep multi-band
near-infrared stack. The Cluster Lensing and Supernova survey
with {\it Hubble} (CLASH; Postman et al. 2012) has discovered
several $z>8.5$ candidates, three at $z\sim 9-10$ (Zheng et al.
2012; Bouwens et al. 2012) and a multiply-imaged source at 
$z=10.7$ (Coe et al. 2013). 

The deepest search to date for star-forming galaxies beyond
a redshift $z\sim 8.5$ utilizing a new sequence of the 
WFC3/IR images of the Hubble Ultra Deep Field (UDF12) was
just completed in September 2012 (Ellis et al. 2012). This
search has thus far yielded 7 promising $z>8.5$ candidates,
including the recovery of UDFj-39546284 (previously proposed
at $z=10.3$, though now suspected of lying at $z=11.8$). It
is not entirely clear whether this object is really a young
galaxy at such a high redshift, or is perhaps an intense emission 
line galaxy at $z\approx 2.4$. Both interpretations are problematic
(see Bouwens et al. 2011; Bouwens et al. 2013; Brammer et al. 2013).

The highest redshift galaxies discovered so far are shown in
Table 1, together with best estimates of their SFRs and current
masses. Galaxies observed with deep-exposure imaging are too faint
to yield masses directly. In these cases, their mass is inferred
by comparing their SFR with the specific star formation rate 
(sSFR$\sim 2-3$ Gyr$^{-1}$) observed on average for galaxies over 
a broad range of redshifts ($z\sim 2-7$; see, e.g., Stark et al. 
2009; Gonzalez et al. 2010 McLure et al. 2011). 

Other than the candidate UDFj-39546284 which may not even be a
high-$z$ galaxy, the most distant galaxy reliably known to date, 
with a photometric redshift of $z\approx 10.7$, is MACS0647-JD
(Coe et al. 2013). Estimates of its mass are based on the
aforementioned sSFR measured at lower redshifts, but also 
on the fact that the average stellar mass ($\sim 10^9\;M_\odot$)
of galaxies at $z\sim 7-8$ rose to a few times $10^{10}\;M_\odot$
by $z\sim 2$ (Gonzalez et al. 2010). Based on this trend, 
one might expect the average stellar mass of galaxies at
$z\sim 11$ to be $\la 10^9\;M_\odot$. If this holds true
for MACS0647-JD, its stellar mass would be on the order
of $10^8-10^9\;M_\odot$. In $\Lambda$CDM, this redshift 
corresponds to a cosmic time $t\sim 427$ Myr. Therefore,
almost a billion solar masses had to assemble inside
this galaxy in only $\sim 130$ Myr.

However, the high-$z$ galaxy with the best measured properties
appears to be MACS1149-JD, a gravitationally lensed source at 
reshift $z\sim 9.7$ (Zheng et al. 2012). The availability
of both {\it Hubble} and {\it Spitzer} data has made it
possible to produce a broad spectrum in the object's rest
frame which, when combined with state-of-the-art models of
synthetic stellar populations, yields reliable estimates
for several key parameters. This analysis suggests a stellar
mass of $\sim 1.5\times 10^8\;M_\odot$ and a SFR of
$\sim 1.2\;M_\odot$ yr$^{-1}$. It is comforting to note
that the estimates of $M$ and SFR for the other high-$z$
galaxies in Table 1 are consistent with these measurements.

{\baselineskip 12pt
\begin{deluxetable*}{lrccl}
\tablewidth{300pt}
\tablecolumns{8}
\tablewidth{0pc}
\tablecaption{High-$z$ Galaxies}
\tablehead{
\colhead{Name} & $ z\;\;$ &$\;\;M$ ($10^8\,M_\odot$)& $\;$SFR ($M_\odot$ yr$^{-1}$)    
&Refs.
}
\startdata
UDFj-39546284$^\dag$&11.8& $\sim10$ &$\sim 4$&1,2,6,7\\
MACS0647-JD&10.7&$1-10$&$1-4$&3\\
MACS1149-JD&9.7&$\sim 1.5$&$\sim 1.2$&4,5\\
UDF12-4106-7304&9.5&$\sim 10$&$\sim 3$&6\\
UDF12-4265-7049&9.5&$\sim 10$&$\sim 3$&6\\
MACSJ1115-JD1&9.2&$\sim 4$&$\sim 1$&5\\
MACSJ1720-JD1&9.0&$\sim 4$&$\sim 1$&5\\
UDF12-3921-6322&8.8&$\sim 10$&$\sim 3$&6\\
UDF12-4344-6547&8.8&$\sim 10$&$\sim 3$&6\\
UDF12-3895-7114&8.6&$\sim 10$&$\sim 3$&6\\
UDF12-3947-8076&8.6&$\sim 10$&$\sim 3$&6
\enddata
\tablenotetext{References:\hskip0.2in} {(1) Bouwens et al. (2011);
(2) Oesch et al. (2013); (3) Coe et al. (2013); (4) Zheng et al. (2012);
(5) Bouwens et al. (2012); (6) Ellis et al. (2013); (7) Brammer et al. (2013) \\
$^\dag$This source may instead be an evolved galaxy at $z\sim2.4$, though both
interpretations are problematic. \\
 \\
}
\end{deluxetable*}
}

\subsection{Inferred Evolutionary History in $\Lambda$CDM}
The comparison between theory and observations shows that 
the calculations can account quite well for
the empirically derived average SFR based on a comprehensive 
analysis of $z\la 8$ galaxies using the ultra-deep WFC3/IR data (Oesch et al. 
2013). These data sets exploit all of the WFC3/IR observations from HUDF09, 
HUDF12, as well as wider area WFC3/IR imaging over GOODS-South. (This 
compilation also produced three samples centered around $z\sim 9$, $z\sim 10$, 
and $z\sim 11$, with seven $z\sim 9$ galaxy candidates, and one each at 
$z\sim 10$ and $z\sim 11$.)

At least in this redshift range, the simulations thus appear to
be handling the galaxy evolution rather well. It is therefore problematic
that none of the calculations completed thus far have shown the possibility
of forming $10^8-10^9\;M_\odot$ galaxies prior to redshift $\sim 10$. 
To better understand why the basic physical principles summarized in
\S~2.1 produce this outcome, we will take the following approach.
We will assume that a galaxy of mass $M\sim 10^9\;M_\odot$ at $z\sim 10$ 
would have followed a growth history similar to a comparably massive galaxy 
at $z\la 8$, though with an appropriate translation in cosmic time $t$. 
This seems reasonable because $t$ also functions as the proper time in 
the local co-moving frame, so that locally (at least) the growth rate 
would not be overly affected by the universal expansion rate, i.e., by 
the functional form of $z(t)$. (This cannot be completely true, however, 
since the expansion history does affect the rate at which a local condensation 
of mass-energy can accrete from larger volumes. Our analysis here should 
therefore be viewed as preliminary, with the need of a more comprehensive 
simulation to follow before our conclusions can me made definitive.)

Let us now consider the evolutionary growth of the galaxies in Table~1 
as a function of cosmic time. Figure~1 shows the history of three
representative members of this set, calculated using the analytic
(exponential) form of the time-dependent SFR from Jaacks et al. (2012),
though with the time $t_*$ shifted to coincide with the measured redshift.
We include the candidate UDFj-39546284, though there are concerns about
whether this source is truly a galaxy at redshift $z\sim 11.8$, rather
than an interloper at $z\sim 2.4$. However, our conclusions are not
affected either way, though if this is indeed a high-$z$ galaxy, the
conflict with $\Lambda$CDM is even worse than that implied by the
other galaxies on this list.

Figure~1 also shows several critical periods in the early Universe. These
include (i) the Dark Ages at $t\sim 0.4-400$ Myr, (ii) the Epoch of
Reionization at $t\sim 400-900$ Myr, (iii) the time at which Population
III stars created the necessary conditions for the subsequent formation
of Population II stars (at $t\sim 300$ Myr), and (iv) the time during
which $5-10\;M_\odot$ black-hole seeds would have had to form in
$\Lambda$CDM in order for them to grow via Eddington-limited accretion
into the $\sim 10^9\;M_\odot$ quasars seen at $z\sim 6-7$ (Melia 2013a).

\begin{figure}[hp]
{\centerline{\epsscale{1.0} \plotone{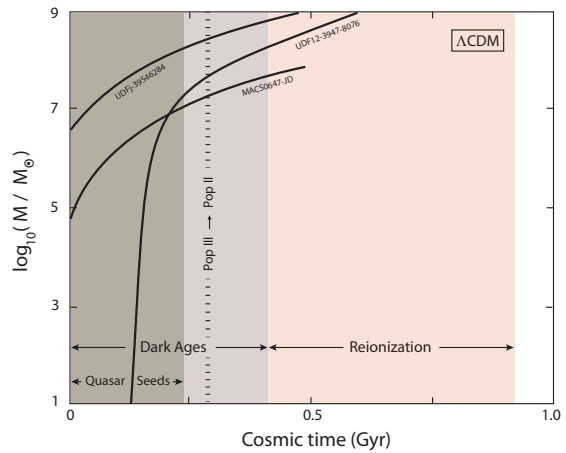} }}
\figcaption{Growth of high-$z$ galaxies observed at $z\sim 9-11$, as a function
of cosmic time $t$, in $\Lambda$CDM. The principal epochs are (i) the ``Dark Ages" 
($t\sim 0.4-400$ Myr), (ii) the period of reionization ($t\sim 400-900$ Myr), (iii)
the transition from Population III to Population II star formation (at $\sim 300$ Myr)
and (iv) the era ($t\sim 0-240$ Myr) during which $5-20\;M_\odot$ seed black holes 
would have had to form in order to grow (via Eddington-limited accretion) into the 
high-$z$ quasars observed at $z\sim 6-7$ (see Melia 2013a). The galaxy growth rate 
is based primarily on the observed SFR at $z\sim 9-12$ (Oesch et al. 2013), 
supported by the theoretical simulations discussed in \S~2. For the sake of
clarity, only 3 representative galaxies from the compilation in Table 1 are
shown here. The others have very similar evolutionary histories.}
\end{figure}

Our earlier conclusion that $\Lambda$CDM would not have allowed the
Universe sufficient time to form high-$z$ quasars consistent with our
current understanding of how structure evolved during the first
$\sim 900$ Myr is strongly reinforced by very similar, and compatible,
issues regarding the premature formation of high-$z$ galaxies. In some
ways, the conflict between theory and observations is worse for the
latter because the physics of Population II star formation, and their
subsequent assembly into primordial galaxies, does not allow enough
flexibility for $\sim 10^9\;M_\odot$ structures to appear in only 
$\sim 200$ Myr. As this figure shows, some of the high-$z$ galaxies
would have had to start forming right after the big bang; several
would have had to start with a mass $M\ga 10^5-10^6\;M_\odot$ {\it right 
at} the big bang. Clearly, there simply wasn't enough time in 
$\Lambda$CDM for the first galaxies to form by redshift $z\sim 
10-11$.  

\subsection{Inferred Evolutionary History in the $R_{\rm h}=ct$ Universe}
The tension between theory and observations disappears in the $R_{\rm h}=ct$
Universe. The $R_{\rm h}=ct$ Universe is an FRW cosmology whose basic
principles follow directly from a strict adherence to the Cosmological
Principle and Weyl's postulate (Melia 2007; Melia \& Shevchuk 2012; see also
Melia 2012a for a pedagogical treatment). Over the past several years,
the predictions of $R_{\rm h}=ct$ have been compared with those of
$\Lambda$CDM and with the available observations, both at low and high 
redshifts. These include cosmic chronometer data, which appear to favor 
the expansion history in the former (Melia \& Maier 2013), and the 
gamma-ray burst Hubble diagram, which confirms that $R_{\rm h}=ct$ 
is more likely to be correct than $\Lambda$CDM (Wei et al. 2013). 
A short summary of the current status of this cosmology appears 
in Melia (2012b, 2013b).

\begin{figure}[hp]
{\centerline{\epsscale{1.0} \plotone{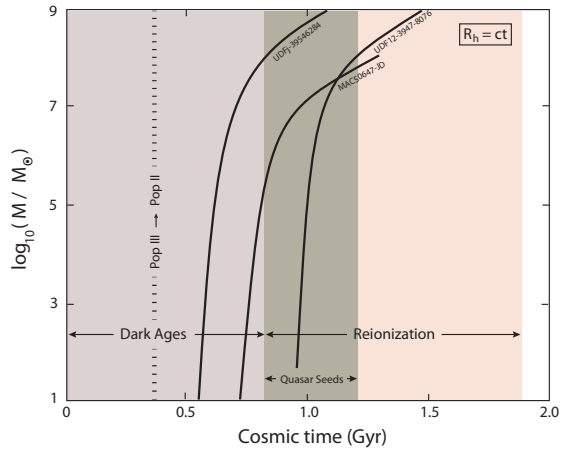} }}
\figcaption{Same as figure~1, except here for the $R_{\rm h}=ct$ Universe. In this case,
the EoR corresponds to $t\sim 800-1,900$ Myr, and the Dark Ages extend up to $\sim 800$ Myr.
The black-hole seed formation period is now $t\sim 800-1,200$ Myr, well within the EoR.
Note, in particular, that all of the high-$z$ galaxies observed thus far would have
started their growth {\it after} the transition from Population III to Population II
star formation at $\sim 300$ Myr.}
\end{figure}

Insofar as accounting for the high-$z$ quasars and galaxies is concerned, 
the essential feature of the $R_{\rm h}=ct$ Universe that distinguishes it 
from $\Lambda$CDM is its expansion factor, $a(t)\propto t$. In this cosmology,
the redshift is given as
\begin{equation}
1+z={t_0\over t}\;,
\end{equation}
where $t_0$ is the current age of the Universe and $t$ is the cosmic time at which
the light with redshift $z$ was emitted. In addition, the gravitational horizon $R_{\rm h}$
is equivalent to the Hubble radius $c/H(t)$, and therefore one has $t_0={1/ H_0}$.
With these equations, we can produce a diagram like that shown in figure~1,
except this time for the $R_{\rm h}=ct$ Universe, and this is illustrated in 
figure~2. The EoR redshift range $z\sim 6-15$ here corresponds to the cosmic time 
$t\sim 830- 1,890$ Myr, so the Dark Ages did not end until $\sim 830$ Myr after 
the big bang. Note also that in this cosmology, $5-20\;M_\odot$ black-hole
seeds formed during the period $t\sim 830-1,200$ Myr, at the beginning of the
EoR, would have grown via Eddington-limited accretion into the $\sim 10^9\;M_\odot$
quasars seen at $z\sim 6-7$. 

Our examination of the evolutionary history of high-$z$ galaxies confirms and
reinforces the view that $R_{\rm h}=ct$ may be better than $\Lambda$CDM in accounting 
for the timing and duration of various important epochs in the early Universe.
From figure~2, we see that in $R_{\rm h}=ct$, not only did all the high-$z$
galaxies have time to grow into $\sim 10^9\;M_\odot$ structures by redshift
$z\sim 10-11$, but all of them (including UDFj-39546284) would have done so
{\it after} the transition from Population III to Population II star formation
at $t\sim 300$ Myr. This would appear to be a minimal requirement in any successful
cosmological theory, given that the assembly of primordial galaxies could not
have started until after the clouds that would make Population II stars had
begun their condensation. 

But how accurately do we know the masses of these galaxies, and is it
possible that our current estimates of $M$ are simply greatly overestimated?
To gauge the possibility that the timing problem in $\Lambda$CDM may be
due to such uncertainties, we show in figure~3 the impact of changing
$M$ by a factor 10 in two representative galaxies from the sample listed
in Table 1. These two objects span the range of redshifts covered in this
group, other than UDFj-39546284 which, as we have already noted, may 
simply be a mis-identification. The shaded regions in this figure show the
possible evolutionary trajectories of these galaxies in both $\Lambda$CDM
and $R_{\rm h}=ct$, when we allow for an order-of-magnitude uncertainty
in the mass. It is quite straightforward to see that the problem still persists.
In order to alleviate the negative impact of the time compression in the
standard model, the inferred masses for these objects would have to
be wrong by at least 4 or 5 orders of magnitude. Even with the range of
masses used in this figure, none of the galaxies in this sample could
have grown to their observed size following the transition from 
Population III to Population II stars, and at least one member of 
this set would have had to start growing before the big bang.

Finally, one of our stated goals at the beginning of this discussion was to 
see if the creation of $5-20\;M_\odot$ black-hole seeds could be made
compatible with the evolutionary growth of structure in the early Universe.
Figure~2 suggests that the answer is yes. If the high-$z$ quasars grew at
an Eddington-limited accretion rate, they would have started as black holes
created in Population III (and possibly Population II) stars near the
beginning of the EoR. In fact, this diagram even allows for the possibility
that their growth was slower than the standard Eddington rate, since
Population III stellar explosions could have occurred up to $\sim 500$ Myr
earlier. According to this figure, most of the primordial galaxy growth
also took place within a period of roughly $600$ Myr, starting near the
end of the Dark Ages, and extending into the middle of the EoR. Quite
tellingly, most of this growth also would have coincided with the formation
of the $5-20\;M_\odot$ black-hole seeds.

However, it is still too early to tell whether the black hole seeds formed
during Population III supernovae, or later, from the explosions of Population
II stars. Perhaps the answer is both. Even if a Population III supernova would
have produced the seed, the fact that the explosion expelled most of the
material, forcing a delay by another $\sim 100$ Myr before gravitational 
collapse could have replenished the gaseous medium surrounding the black
hole, would have significantly slowed down the rate at which the quasar
could have grown initially. It is likely that regardless of whether the black hole
seeds formed at $\sim 300-400$ Myr, or later within primordial galaxies
(at $t\sim 1$ Gyr), their rapid growth into $\sim 10^9\;M_\odot$
supermassive black holes would have been delayed until $t\sim 1-1.9$ Gyr.
But this kind of timing flexbility is only available in $R_{\rm h}=ct$.
It is not possible in $\Lambda$CDM

\section{Conclusions}
The problem with the premature formation of supermassive black holes
at $z\ga 6$ has been with us for several years. In order to get around
the limited time available in $\Lambda$CDM to form such enormous 
objects in only $\sim 500$ Myr, various `fixes' have been proposed
and studied, though the latest observations suggest that even these
modifications may not be consistent with the data. For example, 
the possibility that black holes might have grown at greatly
super-Eddington rates seems to have been ruled out by the
latest measurements (e.g., Mortlock et al. 2011; De Rosa et al. 2011;
Willott et al. 2012), which indicate that the most distant quasars are 
accreting at no more than the standard Eddington value. The possibility
that their seeds may have been much more massive than is typically
seen ($5-20\;M_\odot$) in supernova explosions appears
to be remote, at best, given that we simply don't see these forming
in the local Universe. At the very least, new physics would have to be
devised in order to account for such exotic events.  All in all, our
ever improving capability to constrain the high-$z$ quasar properties
appears to be arguing against the viability of $\Lambda$CDM to account
for the expansion of the early Universe.

The dramatic recent discovery of high-$z$ galaxies has created
renewed interest in this timing problem, because these structures
appear to have formed too quickly following the end of the
Dark Ages.  In some instances, $\sim 10^9\;M_\odot$ of
stellar material would have necessarily assembled in only 
$\sim 200$ Myr. This is inconsistent with the broad range
of detailed simulations carried out to date. The premature
emergence of high-$z$ galaxies confirms and reinforces
the difficulty faced by $\Lambda$CDM in properly accounting
for the formation of structure during the cosmic dawn.

\begin{figure}[hp]
{\centerline{\epsscale{1.0} \plotone{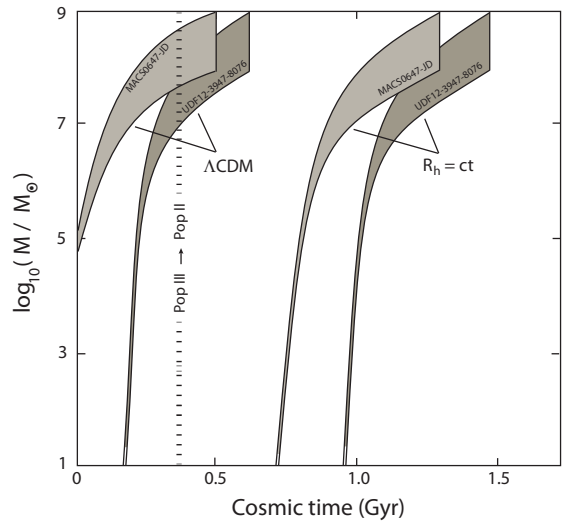} }}
\figcaption{Same as figures~1 and 2, except here highlighting the dependence
of our results on the uncertainty in the inferred galaxy mass $M$. The shaded
regions correspond to a mass range $10^8-10^9\;M_\odot$, the largest uncertainty
quoted in Table 1 (and references cited therein), for two representative objects
from this sample. The transition from Population III to Population II star formation
would have occurred at $\sim 300$ Myr in both cosmologies, but the time
interval corresponding to the redshift range of the EoR is different, so we
do not show it here.}
\end{figure}

In this paper, we have explored the possibility that the
evolutionary growth of high-$z$ quasars and galaxies
might instead be better explained in the context of the
$R_{\rm h}=ct$ Universe. We have found that not only
would it have been possible for these structures to
grow since the big bang consistent with standard physical 
principles, but that their evolution would have been mutually 
consistent with each other, and with what we currently
believe was the history of early star formation. 

Having said this, there are at least two important caveats
to this conclusion. The first clearly has to do with how accurately
we know the mass of these high-redshift galaxies. This question
was partially addressed in figure~3, which showed that even a
factor 10 uncertainty in $M$ is insufficient to alleviate the timing
problem in $\Lambda$CDM. We estimate that in order to 
remove the problem completely, the real mass of these objects
would have to be at least 4 orders of magnitude smaller than
currently measured. Whether this is feasible remains to be 
seen. The second caveat is that although none of the simulations
carried out to date within the context of $\Lambda$CDM can
adequately account for these early galaxies, it may be possible
that a critical physical ingredient has been overlooked. With
the high-$z$ quasars, proposals to fix the problem have 
invoked new physics to generate massive seeds, or unusual
circumstances to permit super-Eddington accretion. It would
be more difficult to generate such ideas for the early formation
of galaxies, which are aggregates of many stars, not single
objects and, at least as far as we know today, could not
have started forming until Population III stars gave way
to Population II. Perhaps the initial cooling that led to the
formation of Population III stars was not due to the 
condensation of molecular hydrogen; maybe some other
process permitted the gas to cool much faster than the
currently believed several hundred Myr time frame. No
doubt, future simulations will probe such ideas and new
physics, and perhaps a solution may be found to work
with the timeline afforded by $\Lambda$CDM. But with
what we know today, our results demonstrate that the
formation of high-redshift galaxies could be consistent
with $R{\rm h}=ct$, but probably not with $\Lambda$CDM.

Our understanding of this important early period in the
Universe's life will improve quite rapidly in the coming
years as primordial galaxies are observed with increasing
numbers and greater sensitivity, and as simulations incorporate
more detailed physics and new ideas.

\vskip-0.2in
\acknowledgments
I am very grateful to the many workers who spent an extraordinary amount of
effort and time accumulating the data summarized in Table~1, and to the anonymous
referee for helpful suggestions and comments that have resulted in a significant improvement
to the manuscript. I am also grateful to Amherst College for its support through a 
John Woodruff Simpson Lectureship.


\begin{thebibliography}

\bibitem[]{} Abel, T., Bryan, G. L. \& Norman, M. L. 2002, Science, 295, 93
\bibitem[]{} Barkana, R. \& Loeb, A. 2001, Phys. Rep., 349, 125
\bibitem[]{} Bouwens, R. J., Illingworth, J. D., Franx, M. \& Ford, H. 2007,
ApJ, 670, 928
\bibitem[]{} Bouwens, R. J. et al. 2011, Nature, 469, 504
\bibitem[]{} Bouwens, R. J. et al. 2012, ApJ, submitted (arXiv:1211.2230)
\bibitem[]{} Bouwens, R. J. et al. 2013, ApJ Letters, in press (arXiv:1211.3105)
\bibitem[]{} Brammer, G. B. et al. 2013, ApJ Letters, in press (arXiv:1301.0317)
\bibitem[]{} Bromm, V., Coppi, P. S. \& Larson, R. B. 2002, ApJ, 564, 23
\bibitem[]{} Bromm, V., Kudritzki, R. P. \& Loeb, A. 2001, ApJ, 552, 464
\bibitem[]{} Bromm, V. \& Larson, R. B. 2004, ARA\&A, 42, 79
\bibitem[]{} Bromm, V., Yoshida, N., Hernquist, L. \& McKee, C. F. 2009, Nature, 459, 49
\bibitem[]{} Chabrier, G. 2003, PASP, 115, 763
\bibitem[]{} Ciardi, B. \& Ferrara, A. 2005, Space Science Reviews, 116, 625
\bibitem[]{} Coe, D. et al. 2013, ApJ, 762, 32
\bibitem[]{} De Rosa, G., DeCarli, R., Walter, F., Fan, X., Jiang, L., Kurk, J., Pasquali, A. and Rix, H.-W.
2011, ApJ, 739, article id. 56
\bibitem[]{} Ellis, R. S. et al. 2013, ApJ Letters, 763, L7
\bibitem[]{} Galli, D. \& Palla, F. 1998, A\&A, 335, 403
\bibitem[]{} Glover, S. 2005, Space Science Reviews, 117, 445
\bibitem[]{} Gonzalez, V., Labb\'e, I., Bouwens, R. J. et al. 2010, ApJ, 713, 115
\bibitem[]{} Greif, T. H., Johnson, J. L., Bromm, V. \& Klessen, R. S. 2007, ApJ, 670, 1
\bibitem[]{} Greif, T. H. et al. 2012, MNRAS, 424, 399
\bibitem[]{} Haiman, Z., Thoul, A. A. \& Loeb, A. 1996, ApJ, 464, 523
\bibitem[]{} Jaacks, J., Nagamine, K. \& Choi, J.-H. 2012, MNRAS, 427, 403
\bibitem[]{} Jiang, L. et al. 2007, AJ, 134, 1150
\bibitem[]{} Johnson, J. L., Greif, T. H. \& Bromm, V. 2007, ApJ, 665, 85
\bibitem[]{} Komatsu, E. et al. 2009, ApJS, 180, 330
\bibitem[]{} Kroupa, P. 2002, Science, 295, 82
\bibitem[]{} Kurk, J. D., Walter, F., Fan, X., Jian, L., Jester, S. Rix, H.-W. and Riechers, D. A. 2009,
ApJ, 702, 833
\bibitem[]{} McLure, R. J., Dunlop, J. S. de Ravel, L. et al. 2011, MNRAS, 418, 2074
\bibitem[]{} Melia, F. 2007, MNRAS, 382, 1917
\bibitem[]{} Melia, F. 2012a, Australian Physics, 49, 83
\bibitem[]{} Melia, F. 2012b, AJ, 144, 110
\bibitem[]{} Melia, F. 2013a, ApJ, 764, 72
\bibitem[]{} Melia, F. 2013b, A\&A, 553, id A76 
\bibitem[]{} Melia, F. \& Maier, R. S. 2013, MNRAS, 432, 2669
\bibitem[]{} Melia, F. \& Shevchuk, A. 2012, MNRAS, 419, 2579
\bibitem[]{} Miralda-Escud\'e, J. 2003, Science, 300, 1904
\bibitem[]{} Mortlock, D. J., Warren, S. J., Venemans, B. P. et al. 2011, Nature, 474, 616
\bibitem[]{} Oesch, P. A. et al. 2013, ApJ, submitted (arXiv:1301.6162)
\bibitem[]{} Omukai, K. \& Nishi, R. 1998, ApJ, 508, 141
\bibitem[]{} Postman, M. Coe, D., Ben\'itez, N. et al. 2012, ApJS, 199, 25
\bibitem[]{} Salvaterra, R., Ferrara, A. \& Dayal, P. 2011, MNRAS, 414, 847
\bibitem[]{} Salvaterra, R., Maio, U., Ciardi, B. \& Campisi, M. A. 2013, MNRAS, 429, 4718
\bibitem[]{} Schaerer, D. 2002, A\&A, 382, 28
\bibitem[]{} Stark, D. P., Bunker, A. J., Ellis, R. S., Eyles, L. P. \& Lacy, M.
2007, ApJ, 659, 84
\bibitem[]{} Stark, D. P., Ellis, R. S., Bunker, A. et al. 2009, ApJ, 697, 1493
\bibitem[]{} Tegmark, M., Silk, J., Rees, M. J., Blanchard, A., Abel, T. \& Palla, F.
1997, ApJ, 474, 1
\bibitem[]{} Wei, J.-J., Wu, Xuefeng \& Melia, F. 2013, ApJ, 772, id.43 
\bibitem[]{} Willott, C. J. et al. 2010, AJ, 140, 546
\bibitem[]{} Wise, J. H. \& Abel, T. 2007, ApJ, 665, 899
\bibitem[]{} Wise, J. H. \& Abel, T. 2008, ApJ, 685, 40
\bibitem[]{} Yoshida, N., Bromm, V. \& Hernquist, L. 2004, ApJ, 605, 579
\bibitem[]{} Yoshida, N., Hosokawa, T. \& Omukai, K. 2012, Prog. Theor. Exp. Phys., 2012, id.01A305
\bibitem[]{} Yoshida, N., Omukai, K. \& Hernquist, L. 2008, Science, 321, 669
\bibitem[]{} Zaroubi, S. 2012, in ``The First Galaxies---Theoretical Predictions and 
Observational Clues," eds. T. Wiklind, B. Mobasher and V. Bromm (Springer), in press (arXiv:1206.0267)
\bibitem[]{} Zheng, W. et al. 2012, Nature, 489, 406

\end{thebibliography}
\end{document}